\documentclass[aps,prl,superscriptaddress,twocolumn,citeautoscript]{revtex4-1}
\usepackage{graphicx}
\usepackage{amsfonts}
\usepackage{amsmath}
\usepackage{amssymb}
\usepackage{bm}
\usepackage{textcomp}
\usepackage{verbatim}
\usepackage{xspace}
\usepackage{soul}
\usepackage{bbold}
\usepackage{mathtools}
\usepackage{dcolumn}
\usepackage{hyperref}
\usepackage[usenames]{color}
\usepackage{subfig}
\usepackage{physics}
\usepackage{units}
\usepackage[export]{adjustbox}
\usepackage{tablefootnote}
\usepackage{lipsum}
\usepackage[dvipsnames]{xcolor}
\usepackage{ragged2e}
\usepackage{booktabs}
\DeclareCaptionJustification{justified}{\justifying}
\def\beq{\begin{equation}}
\def\eeq{\end{equation}}

\newcommand{\blue}[1]{\textcolor[rgb]{0.0, 0.0, 0.0}{#1}}

\newcommand{\TO}{TiO$_2$}
\newcommand{\TOF}{TiO$_{1.99}$F$_{0.01}$}
\newcommand{\MO}{MgO}
\newcommand{\h}{h$^+$}
\newcommand{\e}{e$^-$}
\DeclareMathOperator{\argmax}{arg\,max}

\begin{document}
\title{Machine Learning Small Polaron Dynamics}

\author{Viktor C. Birschitzky$^*$}
\address{University of Vienna, Faculty of Physics and Center for Computational Materials Science, Vienna, Austria}

\author{Luca Leoni$^*$}
\address{Department of Physics and Astronomy "Augusto Righi", Alma Mater Studiorum - Universit\`a di Bologna, Bologna, 40127 Italy}

\author{Michele Reticcioli}
\address{University of Vienna, Faculty of Physics and Center for Computational Materials Science, Vienna, Austria}

\author{Cesare Franchini}
\address{University of Vienna, Faculty of Physics and Center for Computational Materials Science, Vienna, Austria}
\address{Department of Physics and Astronomy "Augusto Righi", Alma Mater Studiorum - Universit\`a di Bologna, Bologna, 40127 Italy}

\begin{abstract}
Polarons are crucial for charge transport in semiconductors, significantly impacting material properties and device performance. 
The dynamics of small polarons can be investigated using
first-principles molecular dynamics (FPMD).
However, the limited timescale of these simulations presents a challenge for adequately sampling infrequent polaron hopping events. Here, we introduce a message-passing neural network \blue{combined with FPMD within the Born-Oppenheimer approximation,} that learns the polaronic potential energy surface by encoding the polaronic state, allowing for simulations of polaron hopping dynamics at the nanosecond scale.
By leveraging the statistical significance of the long timescale, our framework can accurately estimate polaron (anisotropic) mobilities and activation barriers in prototypical polaronic oxides across different scenarios (hole polarons in rocksalt \MO\ and electron polarons in pristine and F-doped rutile \TO) \blue{within experimentally measured ranges.}
\end{abstract}

\maketitle

 {\it Introduction.} 
A small polaron is a spatially localized quasiparticle that forms in polarizable materials due to the attraction of an excess charge carrier to its own self-induced lattice distortion.~\cite{LANDAU1933,Pekar46,Alexandrov2010book,Shluger1993,franchini_polarons_2021}.
When thermally activated, small polarons can travel through the crystal, \blue{constituting one of the primary charge transport mechanisms. In addition, they play} a crucial role in various physical and technologically significant processes, such as optical excitations~\cite{Varley2012,baskurt2024,Tanner2021}, (photo)catalysis~\cite{DiValentin2009, reticcioli_interplay_2019, https://doi.org/10.1002/advs.202305139,doi:10.1021/acsami.4c03941}, photovoltaic applications~\cite{Zhang2023} and device functioning~\cite{LUONG2022}.
Several theories have been developed to understand and predict small polaron mobility~\cite{franchini_polarons_2021} including Marcus-Emin-Holstein-Austin-Mott (MEHAM) theory~\cite{Austin2006, MARCUS1985265, EMIN1969439, Holstein2, Marcus1993}, density functional theory (DFT)~\cite{deskins_electron_2007, kowalski_charge_2010}, Diagrammatic Monte Carlo~\cite{Mishchenko2015} and dynamical mean field theory~\cite{Fratini2003}.
\blue{Moreover, alternative approaches have been proposed to study large or intermediate polaron mobility 
~\cite{giannini_quantum_2019,C0CS00198H,chang_intermediate_2022,doi:10.1021/acsenergylett.1c00506, Frost2017, Zhou2019, Miladic2023,franchini_polarons_2021}.}

\begin{figure}
    \centering
    \includegraphics[width=0.8\linewidth]{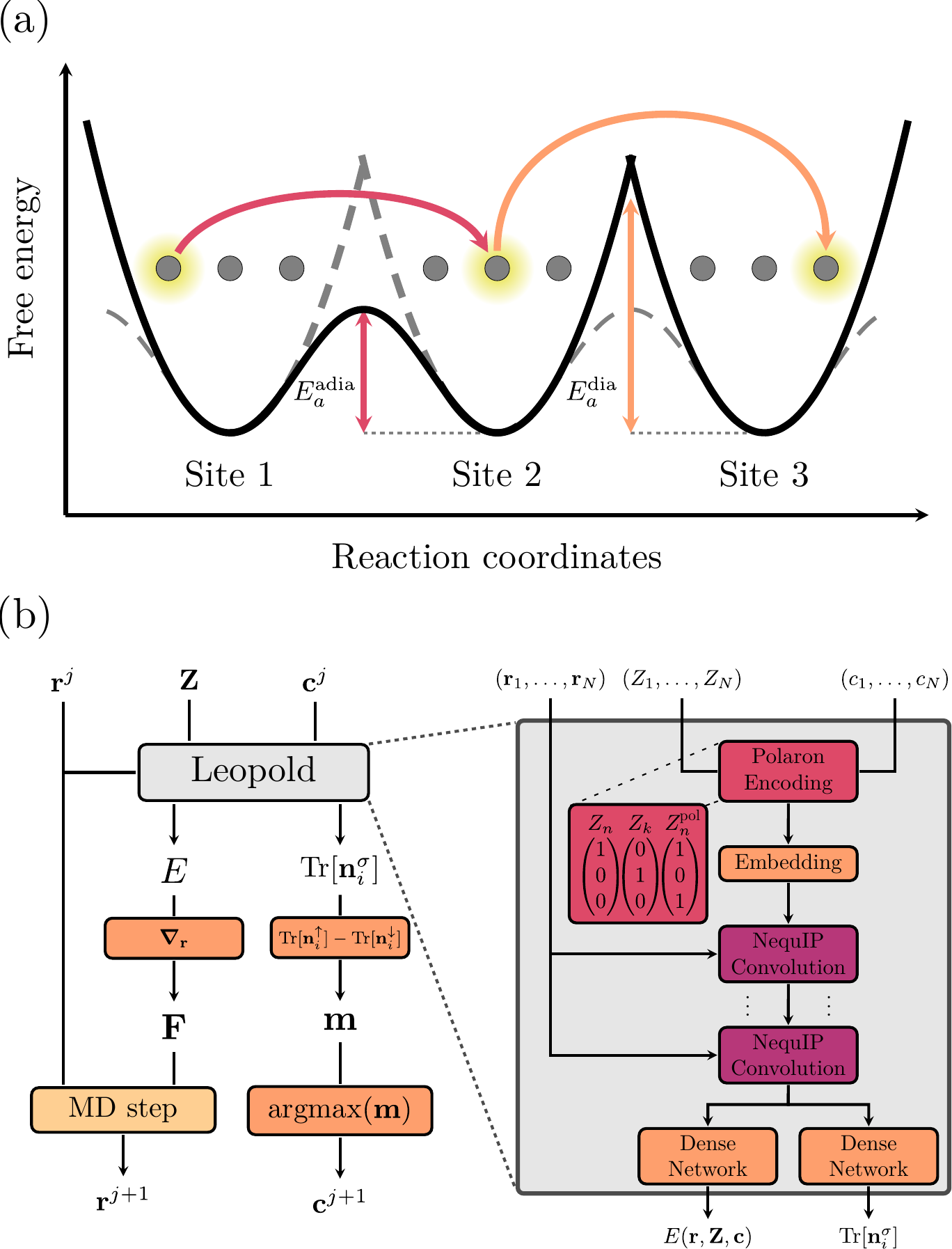}
    \caption{
    (a) Schematic description of adiabatic and non-adiabatic hopping transitions from one site to its neighbor, with activation energy $E_a^{\mathrm{adia}}$ and $E_a^{\mathrm{dia}}$.
    Small filled circles symbolize lattice sites, whereas the light clouds represent the polaron charge.
    (b) Overview of workflow and model architecture. 
    \blue{The left side shows the applied MLMD workflow, where predicted energies $E$ and forces $\mathbf{F}$ are used to integrate the nuclei's equations of motion, while predicted occupations $\mathrm{Tr}[\mathbf{n}^\sigma_i]$ and derived magnetizations $\mathbf{m}$ determine the charge state $\mathbf{c}^{j+1}$ at the next timestep.
    The right side shows the model architecture of Leopold.}
    Positions $\mathbf{r}^j$, nuclear charges $\mathbf{Z}$, and charge states $\mathbf{c}^j$ are fed into the neural network.
    \blue{The polaron encoding appends the one-hot-encoded nuclear charge $Z_j$ and charge state $c_j$ of each atom to represent the polaronic state of the system (the zoom-in shows possible vectors in a binary polaronic compound, where $Z_n^\mathrm{pol}$ encodes a polaronic atom).}
    }
    \label{fig:architecture}
\end{figure}

Small polaron transfer can occur either via hopping mechanism or through delocalization in the conduction band.~\cite{Coropceanu2007, Ortmann2011, natanzon_evaluation_2020}
As depicted in Fig.~\ref{fig:architecture}(a), thermally activated small polaron hopping is either adiabatic or diabatic, depending on the strength of the electronic coupling between the initial and final states. In the adiabatic regime, the polaron charge transfers smoothly to a neighboring site, while in the diabatic regime, the charge is transferred instantaneously~\cite{MARCUS1985265, Marcus1993}. 
Unlike conduction band diffusion, for which mobility is difficult to estimate~\cite{C4CP03981E}, polaron hopping mobility can be obtained using MEHAM theory and the Einstein relation~\cite{einstein1956}, typically by calculating the activation energy through quasi-static linear interpolation between initial and final states along specific polaron transport paths~\cite{deskins_electron_2007}. 
\blue{While MEHAM-theory allows for a rigorous treatment of the electronic coupling of initial and final states and quantitatively describes both adiabatic and diabatic polaron transfer, the} approach limits the comprehensive exploration of diffusion pathways and introduces approximations for the pre-exponential factor~\cite{natanzon_evaluation_2020}.

The dynamical aspects of polarons can be explored using first principle\blue{s} molecular dynamics (FPMD), \blue{which is typically based on DFT within the Born-Oppenheimer (BO) approximation.}
FPMD has proven highly successful in assessing dynamical properties of materials~\cite{car_unified_1985} and has been extensively used to study small polarons, offering insights into their activation energies, hopping dynamics, configurational space and charge recombination~\cite{kowalski_charge_2010, setvin_direct_2014, Hao2015, reticcioli_polaron-driven_2017, reticcioli_formation_2018, Zhang2019, Wiktor2018,Selcuk2016,cheng_photoinduced_2022}.
\blue{
While a rigorous treatment of diabatic electron transfer (i.e., tunneling) requires methods beyond the BO approximation~\cite{Xu2022, Xu2023, Wang2023, deskins_electron_2007}, FPMD remains a powerful tool for accurately describing adiabatic electron transfer.
FPMD can qualitatively capture diabatic transfer (hopping from one polaronic potential energy surface to another) when the transfer occurs rapidly enough~\cite{Stuchebrukhov2016}.
However, despite its renowned computational efficiency, FPMD is limited by the accessible timescale, typically spanning only a few picoseconds.}

Since the pioneering works on neural network potentials~\cite{behler_generalized_2007} and Gaussian approximation potentials~\cite{bartok_gaussian_2010}, machine learning (ML) has transformed FPMD~\cite{friederich_machine-learned_2021, unke_machine_2021}.
Contemporary ML-based potential energy surface (PES) 
models often employ message-passing neural networks (MPNN) and incorporate equivariance into their architecture~\cite{batatia_mace_2022,batzner_e3-equivariant_2022}, enhancing both accuracy and data efficiency in learning the PES.
These models enable the assessment of material properties at a fraction of the cost of direct DFT evaluation, extending simulation times or system sizes while preserving DFT accuracy. 

Enhancing the capability of ML-based MD (MLMD) to account for polaron dynamics is complex and remains an open challenge.
While effects such as the prediction of local magnetic moments~\cite{eckhoff_predicting_2020} and non-local charge transfer~\cite{deng_chgnet_2023} have been incorporated into ML-based force fields using local environment descriptors, these methods do not explicitly enforce charge conservation, rendering them unsuitable for simulating small polaron dynamics. To address charge conservation, global charge equilibration schemes have been proposed that allocate charge based on the electronegativity at each site~\cite{ko_fourth-generation_2021,doi:10.1021/acs.jctc.2c01146}. However, these approaches may lead to unphysical charge transport mechanisms due to their global nature and the potential for spurious responses to thermal fluctuations in local environments.

In this letter, we present the ML package Leopold (LEarning Of POLaron Dynamics), designed to accurately model small polaron dynamics. The architecture, implemented in JAX~\cite{jax2018github}, is 
built on a modified version of the Neural Equivariant Interatomic Potential (NequIP) architecture~\cite{batzner_e3-equivariant_2022} trained on Vienna Ab initio Simulation Package (VASP) data~\cite{PhysRevB.54.11169,KRESSE199615}. 
We introduce two key features to account for polaron transport: (\textit{i}) explicit charge state encoding~\cite{zubatyuk_teaching_2021} to ensure charge conservation and identify polaronic sites, and (\textit{ii}) direct prediction of site occupation to track polaron hopping and regularize the descriptor~\cite{deng_chgnet_2023}. 
The model is assessed by considering three different systems: hole polaron in rocksalt \MO~(\MO+\h in oxygen $p$ orbitals), electron polaron in rutile \TO~(\TO+\e in titanium $d$ orbitals), and combined electron polaron-defect dynamics in F-doped \TO~(\TOF). To describe the coupled dopant-polaron system, we employ an active learning scheme that effectively addresses the configurational sampling problem and accurately reproduces the correct polaron distribution in a defective system (see SM for details on the employed methodology~\cite{SM}).
Our approach can predict polaron hopping events, thereby enhancing the sampling of hopping trajectories and enabling nanosecond-scale simulations to extract polaronic mobility, activation energies ($E_\mathrm{a}$), and transition pathways.   

{\it Model.} 
Most ML models, including MPNNs, distinguish atoms based on their chemical species using a one-hot encoded vector of the nuclear charge $Z_i$ but do not account for oxidation states. 
Since small polarons discretely alter the oxidation state by ±1, we \blue{can encode each atom’s charge state in the input one-hot-encoded vector of Leopold.
    The charge state $\mathbf{c}$ is determined through the following relation:
    \begin{equation}
        c_i = 
        \begin{cases}
            1, & \text{if $i=\argmax_j |m_j|$}\\
            0, & \text{otherwise}
        \end{cases}
    \end{equation}
    where $c_i$ is the charge state of atom $i$ ($c_i=1$ for polaronic site, 0 for non-polaronic sites) and $m_j$ is the magnetization of atom $j$. 
    In the "polaron encoding" (as shown in Figure \ref{fig:architecture}b), the value $c_i$ is appended to the one-hot encoded vector representing the species of atom $i$.
    The resulting encoding vectors in a polaronic binary compound is a three-dimensional vector (e.g., in \TO : Ti$=(1,0,0)$, O$=(0,1,0)$, Ti$^\mathrm{pol}=(1,0,1)$).
    The description can easily be generalized to systems with $n$ polarons by assigning $c_i=1$ to the $n$ maximally magnetized atoms.}
    
    Polarons can hop to adjacent sites when more favorable local bonding geometries enhance the adiabatic electronic coupling.
To capture these dynamics, which are not reflected in the discrete encoding of charge states, Leopold learns the local charge occupation of each site through the architecture reported in Fig.~\ref{fig:architecture}(b).
This approach serves two key purposes: it enables tracking and updating of the polaronic states, while also regularizing the learned representation to incorporate polaronic state information.
This is done by learning the occupation matrix $\mathbf{n}$, defined as:
\begin{equation}
    n_{lmm'}^{\sigma} = \sum\limits_i f_i \bra{\psi_i^\sigma} P_{lmm'}^{\sigma} \ket{\psi_i^\sigma}
\end{equation}
where $\psi_i$ are the Kohn-Sham states, $f_i$ their occupations, $l$ the atom index, $m$ the specific projector function, and $\sigma$ the spin channel. 
We use $p$-state-projectors for \MO+\h\ and $d$-state-projectors for pristine and F-doped \TO, as defined in DFT+\textit{U}~\cite{PhysRevB.57.1505}. 
We checked that, instead of predicting the full occupation matrix, it is sufficient to learn the invariant atomic quantity $\mathrm{Tr}[\mathbf{n}^\sigma_i]$ for each atom $i$ and spin channel $\sigma$, denoted as $n^\sigma_i$. 
Since small polarons carry a spin, to stabilize the prediction of the charge occupations, we impose constraints ensuring that the total occupation 
$n^\uparrow_i + n^\downarrow_i$, the magnetization $n^\uparrow_i - n^\downarrow_i=m_i$, and the 
total magnetization of the system $\sum_i n^\uparrow_i - n^\downarrow_i$ 
are also fitted to the reference data. 
For details on the machine learning methodology see the SM~\cite{SM}.

\begin{figure*}
    \centering
    \includegraphics[width=0.8\linewidth]{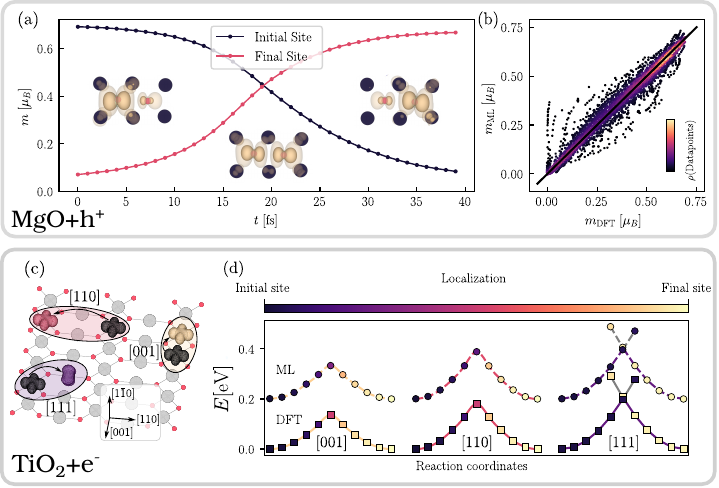}
    \caption{
    Polaron hopping from ML and DFT for hole polarons in MgO and electron polarons in TiO$_2$. 
    (a) Hopping event extracted from MLMD through the evolution of the local site-projected magnetization $m$. Polaron charge isosurfaces are shown for the initial, intermediate (crossing point), and final states. 
    (b) Comparison of $m_\mathrm{DFT}$ and $m_\mathrm{ML}$ of 50 hopping events extracted from MLMD at various temperatures. $m$ is only displayed for the involved sites ($m>0.3$ at any timestep in the hopping event) and the data point density is encoded via the color scale. 
    (c) Possible polaron transitions to neighboring sites along three different directions ([001], [110], [111]) in \TO\ shown for a single TiO$_2$ (1$\overline{1}$0) layer. 
    (d) Comparison between DFT (full lines, square marker) and ML (dashed lines, round marker) activation energies for the three different transition directions as approximated by linear interpolation.
    The color code indicates the site-projected charge localization defined through the magnetization $m$ as $(m_\mathrm{final} - m_\mathrm{initial} + 1) / 2$ (0=full localization on the initial site, 1=full localization on the final site).
    At DFT-level [001] and [110] pathways show adiabatically coupled transition states (polaron shared between initial and final sites), while [111] transition is characterized by an instantaneous hopping at the crossing point.
    } 
    \label{fig:hopping}
\end{figure*}

{\it Results.} \blue{Fig.~\ref{fig:hopping}(a) displays a hopping event -- as generated by Leopold in an MLMD-simulation and reproduced in DFT -- based on the dynamical magnetization and demagnetization of two neighboring sites in \MO+\h (other site's magnetizations are close to 0 and omitted for clarity)}.
Initially ($t=0$), the polaron charge is localized on one site and becomes progressively destabilized due to thermal motion. 
At the crossing point, the polaron is shared between two sites until it fully localizes on the final site ($t \approx$ 40~fs). 
To accurately reproduce these hopping processes by our ML model, we initially train Leopold on one FPMD run unavoidably including only a few hopping events and concurrently use the trained model to simulate polaron dynamics in an ML-based MD run. 
As hopping is a relatively rare event in the ps time-scale, the initial dataset may not capture all possible transition pathways.
To address this, we employ an active learning scheme by extracting hopping processes from the ML-based run and calculating them at the reference DFT+\textit{U} level. 
    \blue{This extended dataset is then used to retrain the model iteratively.
    While this approach is not necessarily comprehensive, we found that after a few active learning loops, hopping events were well sampled and the model did not produce new symmetrically-inequivalent transition pathways.}

The results 
are presented in Fig.~\ref{fig:hopping}(b), showing the local magnetization of sites involved in hopping events as produced by Leopold and confirmed by DFT for \MO+\h. 
\blue{Achieved errors are within the expected ranges \cite{eckhoff_predicting_2020, deng_chgnet_2023} as detailed in the SM \cite{SM}.}
From approximately 50 hopping events extracted at temperatures ranging from 100~K to 600~K from 10 ns MLMD runs, we identified two key characteristics of the hopping behavior in \MO+\h: 
(\textit{i}) In the rocksalt octahedral symmetry, the oxygen \textit{p}-states hosting the hole polaron are three-fold degenerate, which allows for only one 
symmetrically inequivalent nearest-neighbor
hopping pathway along the [110] direction; 
(\textit{ii}) The polaronic transition between sites occurs adiabatically~\cite{MARCUS1985265} following an hopping mechanism similar to the one shown in Fig.~\ref{fig:hopping}(a).

In contrast, rutile \TO\ exhibits more complex structural symmetries and hopping processes.  Figs.~\ref{fig:hopping}(c-d) summarize the results, where we focus our discussion to the [001], [110] and [111] hoping paths within a single (1$\overline{1}$0) plane for simplicity.
    The electron polaron occupies a single $t_{2g}$ orbital, where the triple degeneracy is lifted by an axial elongation of the octahedron~\cite{yang_fluorine_2010, PhysRevB.87.125201}. 
    This orbital symmetry favors polaron migration along the [001] and [110] pathways. 
    Polaron transport along [111] requires a 90-degree rotation of the polaron orientation, making these transport directions less probable~\cite{deskins_electron_2007}.
    These considerations are well reflected in our MLMD data at high temperatures, showing 74.3\% of transitions along the [001] direction, 25.5\% along the [110] direction, and 0.2\% along the [111] direction. 
    
    Interestingly, unlike \MO, \blue{hops along} both adiabatic and diabatic \blue{transition pathways (as previously determined by MEHAM-theory~\cite{deskins_electron_2007})} are observed. 
    To better describe these different processes, we performed quasi static calculations, i.e., we calculated energy and polaron localization on structures obtained by linear interpolation of atomic coordinates between initial and final states. 
    As shown in Fig.~\ref{fig:hopping}(d) the ML results are in excellent agreement with DFT calculations.
    Transport along the dominant [001] direction is adiabatic, as deduced by the continuous change of the site-projected magnetization during the hopping process 
    for intermediate interpolated structures between nearest neighbour Ti sites (see gradient bar in Fig.~\ref{fig:hopping}(d) \blue{or the SM \cite{SM}}). 
    Transitions along the less favorable [110] direction between third nearest-neighbor Ti sites can still be classified as adiabatic, as indicated by the smooth charge transfer, as depicted by the color coding in  Fig.~\ref{fig:hopping}(d).
    However, the electronic coupling between the initial and final states, which are 6.4~\AA~apart, is low, as suggested by the more pronounced peak in the corresponding energy curve.
    The less frequent migration pathway along the [111] direction, which requires a modification of the polaron orbital symmetry, is characterized by a diabatic process~\cite{deskins_electron_2007}, associated with a sudden change in the magnetic moment and instantaneous hop from the initial and final states.
    
    Remarkably, even though electron delocalization associated with migrations along the [110] and [111] directions is not fully captured by our integer charge state assumption, Leopold is still able to \blue{qualitatively reproduce the relative frequency of adiabatic and diabatic events~\cite{deskins_electron_2007}.}
    In particular, while the hopping events that exhibit some degree of charge delocalization may cause nonphysical oscillations in our ML model, they remain meaningful as long as the final localization site is consistent with DFT predictions and the transition features a well-sampled energetic barrier.
    Furthermore, the small time scale of unphysical oscillations, compared to the large time scales considered, renders the impact on physical observables computed on the obtained polaron trajectory insignificant \blue{(see the SM for further discussion\cite{SM})}

\begin{figure}
    \centering
    \includegraphics[width=0.8\linewidth]{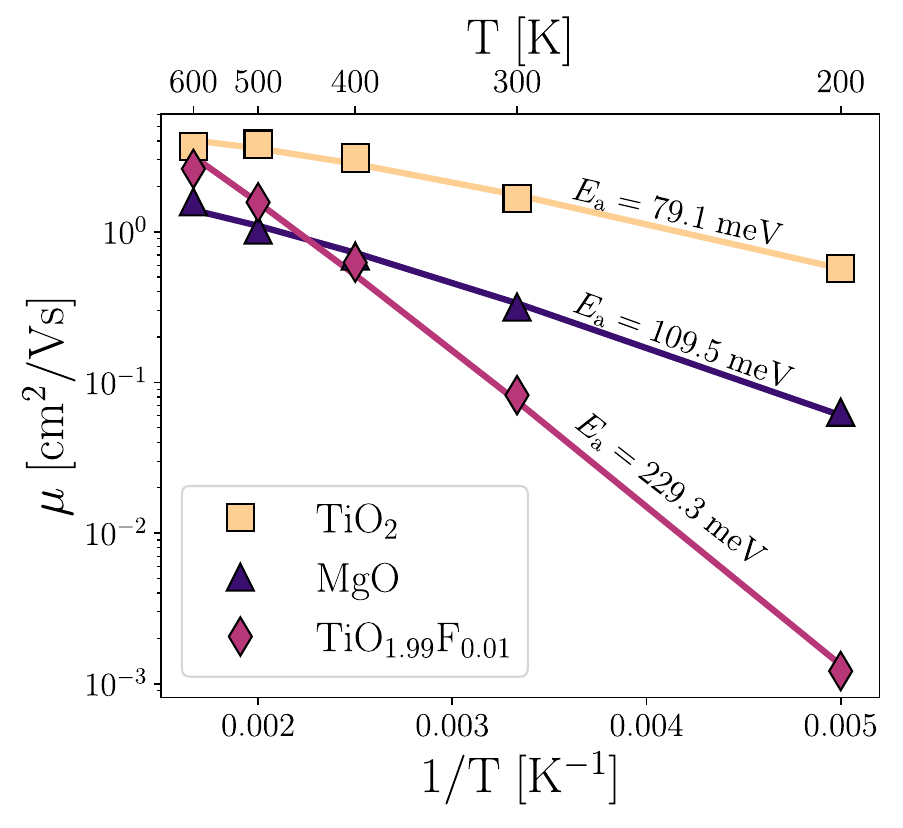}
    \caption{Polaron mobility $\mu$ extracted from the MSD through the  Einstein-relation from ns MLMD NVT-simulations at various temperatures for \TO+\e, \MO+\h and \TOF.}
    \label{fig:Diffusion}
\end{figure}

The polaron transport data generated by Leopold enable an accurate evaluation of diffusion properties.
By using statistical quantities such as the mean squared displacement (MSD) (see SM for details~\cite{SM}), we can evaluate polaron mobility and activation energies, refining the results obtained by quasi-statical models.
Fig.~\ref{fig:Diffusion} collects the mobility computed from NVT simulations at various temperatures in \MO, \TO\ and \TOF\ for up to 10 ns. 
Mobility increases with temperature, as expected for small polaron transport~\cite{franchini_polarons_2021}.
The predicted $\mu$ and $E_\mathrm{a}$ for rutile \TO\ fall within the experimentally reported range of 0.01-10.0~cm$^2$/Vs \cite{morita_models_2023, PhysRevB.54.7945, PhysRevB.69.081101,FUJISHIMA2008515,Tang1994} and agree with previous MEHAM theory work~\cite{deskins_electron_2007}. 
The reduced mobility in \TOF\ is attributed to the attractive interaction between electron polarons and positively charged defects, which typically decreases polaron diffusion. 
Activation energies in \TO+\e\ align with theoretical values nudged elastic band (NEB) calculations ($E_\mathrm{a}=58$~meV using hybrid functional)~\cite{morita_models_2023}.
We note that our model takes into account several types of hopping processes, with different energy barriers, while quasi-static approaches typically consider only one or few processes. 
Near room temperature conductivity measurements for \MO\ reported difficulties in obtaining reproducible results~\cite{doi:10.1021/acs.chemmater.7b00217,ALempicki_1953,KATHREIN1983177}, but activation energies from MLMD simulations agree well with theoretical values based on NEB for \MO+\h~\cite{doi:10.1021/acs.chemmater.7b00217} ($E_\mathrm{a}=110$~meV).

\begin{figure}[t]
    \centering
    \includegraphics[width=0.8\linewidth]{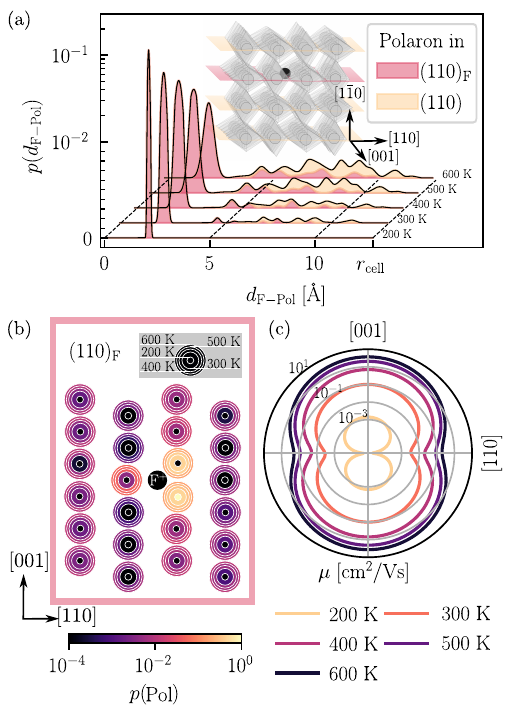}
    \caption{Polaron-Defect-Interaction and charge mobility anisotropy in \TOF\ from 10~ns long MLMD at different temperatures. 
    (a) F-polaron distance distribution; light (yellow) and dark (pink) planes show contributions from different planes highlighting preferential polaron localization in the F-aligned $(110)_\mathrm{F}$-plane and increasing inter-layer mobility with rising temperature.
    (b) Layer-projected site distributions of polaron in $(110)_\mathrm{F}$.
    Innermost circles show distribution at 200~K and consecutively larger circles show distribution at 300, 400, 500, and 600~K.
    (c) Polar plot of the polaron mobility $\mu$ in the $(110)_\mathrm{F}$-plane at different temperatures along [001] and [110] directions.}
    \label{fig:defect}
\end{figure}

We now consider in greater detail a more realistic scenario in which the excess charge forming the polaron is introduced via chemical doping, specifically by substituting an oxygen atom with a fluorine atom in rutile \TO\ (\TOF). In this case, identifying polaronic configurations requires a specialized approach to capture the complex interactions between polarons and defects, and to determine the most favorable polaron-defect coupled configuration.~\cite{birschitzky_machine_2022, birschitzky_machine_2024, doi:10.1021/acs.jpclett.4c00889, zhang2024decouplingmanybodyinteractionsceo2}. 

Polaron trajectory data collected from 10 ns MLMD simulations indicate a strong attractive interaction between the mobile electron polaron and the immobile fluorine impurity, as suggested by the short distance peak in the polaron-fluorine radial distribution function shown in Fig.~\ref{fig:defect}(a). 
The attractive interaction between the electron polaron and the positively charged F defect is evident in the tendency of the polaron to remain near the F defect, restricting its movement within the (1$\overline{1}$0) basal plane containing the F atom. The polaron probability distribution displayed in Fig.~\ref{fig:defect}(b) reveals that the polaron predominantly occupies the Ti sites along the [001] direction adjacent to the F defect. As the temperature increases, the probability of the polaron visiting other sites, including adjacent planes, also increases. However, the short-distance polaron-F complex remains the most favorable configuration.
This result is in good agreement with previous experiments based on electron paramagnetic resonance and electron-nuclear double resonance, which identified the polaronic ground state at low temperatures as the Ti$^{3+}$ ion adjacent to a F$^-$ ion~\cite{yang_fluorine_2010}.

Finally, we show the direction resolved analysis of polaron mobility on (1$\overline{1}$0) plane.
The resulting polar plot displayed in in Fig.~\ref{fig:defect}(c) indicates a strong anisotropy with a preference of the polaron to diffuse in the [001] in agreement with the direction-dependent activation energies in \TO\ shown in Fig~\ref{fig:hopping}(d). This behavior is in very good agreement with bulk conductivity measurement~\cite{Byl2006}.

{\it Conclusions.} To conclude, we developed a machine learning approach to study small polarons dynamics at the nanosecond time scale.
Training our model on polaronic FPMD data reproduces the adiabatic and diabatic polaron transport in two structurally different polaronic materials, \MO\ and \TO, capturing both electron and hole polaron dynamics, as well as polaron-defect correlation.
Our model enables an efficient and comprehensive exploration of polaron hopping pathways, providing substantially increased statistical sampling compared to first-principles MD and offering a dynamic representation of the hopping process. 
Mobilities, activation barriers and diffusion anisotropies are in good qualitative agreement with previous theoretical results and explain measured experimental data. 
The designed ML-aided architecture is general and applicable to other systems including charge transport in organic semiconductors and Li-ion battery materials,
and can be extended to even more complex situations, such as multi-polaron systems, surfaces with adsorbates and exciton-polaron materials.
More efficient generation of training data could eliminate the need for initial FPMD runs, thereby reducing the number of required DFT calculations, allowing for the use of more accurate functionals and further expanding the length scale.
\blue{Furthermore, the generality of the model could be enhanced by incorporating diabatic effects in a more rigorous form by going beyond the BO-approximation.}

\textit{Acknowledgments} -- We thank Dominik Freinberger for useful discussions and initial contributions.
This research was funded by the Austrian Science Fund (FWF) 10.55776/F81 project TACO. C.F. and L.L. acknowledge support by the  
National Recovery and
Resilience Plan (NRRP), Mission 4 Component 2 Investment 1.3 - Project NEST (Network 4 Energy Sustainable Transition) and CN-HPC grant no. (CUP) J33C22001170001, SPOKE 7, of Ministero dell’Università e della Ricerca (MUR), funded by the European Union – NextGenerationEU.
V.B. gratefully acknowledges funding from the Vienna Doctoral School in Physics (VDSP). 
The computational results presented have been achieved using the Vienna Scientific Cluster (VSC).
We acknowledge access to LEONARDO at CINECA, Italy, via an AURELEO (Austrian Users at LEONARDO supercomputer) project.  
$^*$  V.B. and L.L. contributed equally to this work.

\textit{Data availability} — 
The supporting data and code for this article will be made available upon publication.


\nocite{PhysRevB.54.11169,KRESSE199615,PhysRevB.59.1758, PhysRevB.50.17953, PhysRevLett.77.3865, PhysRevB.57.1505, falletta_hubbard_2022, setvin_direct_2014, Nose1984, batzner_e3-equivariant_2022, jax2018github, kingma2017, jaxmd2020}

\bibliography{bib}

\end{document}


\title{Supplementary material for 
'Machine Learning of Small Polaron Dynamics'}

\author{Viktor C. Birschitzky}
\address{University of Vienna, Faculty of Physics and Center for Computational Materials Science, Vienna, Austria}

\author{Luca Leoni}
\address{Department of Physics and Astronomy "Augusto Righi", Alma Mater Studiorum - Universit\`a di Bologna, Bologna, 40127 Italy}

\author{Michele Reticcioli}
\address{University of Vienna, Faculty of Physics and Center for Computational Materials Science, Vienna, Austria}

\author{Cesare Franchini}
\address{University of Vienna, Faculty of Physics and Center for Computational Materials Science, Vienna, Austria}
\address{Department of Physics and Astronomy "Augusto Righi", Alma Mater Studiorum - Universit\`a di Bologna, Bologna, 40127 Italy}

\date{\today}	
\maketitle

\section{Data Generation}
\subsection{Density Functional Theory Methods}
    All computations were performed using the Vienna Ab initio Simulation package (VASP) \cite{PhysRevB.54.11169,KRESSE199615,PhysRevB.59.1758, PhysRevB.50.17953} and using Perdew–Burke– Ernzerhof functionals \cite{PhysRevLett.77.3865} within the rotationally invariant DFT+\textit{U} formalism \cite{PhysRevB.57.1505}.
    The bulk systems taken into consideration are rocksalt \MO+\h, rutile \TO+\e, and F-doped rutile TiO$_2$ with a doping concentration of 1/192.
    In particular, to allow the simulation of polarons, we used a $2\times2\times2$ supercell for MgO (resulting in a supercell with $8.449\times 8.449\times 8.449  \mathrm{\AA}^3$ and 64 atoms) and a $2\times2\times6$ supercell for the other two \TO-systems (resulting in a supercell with $12.973\times 12.973\times 17.724 \mathrm{\AA}^3$ and 288 atoms).
    We used a $\Gamma$-centered $2\times2\times2$ k-points mesh for the former and $\Gamma$ only mesh for the latter materials.
    Spin-polarized calculations were performed for every material and the energy cutoff was set to 300~eV for \TO-systems and 500~eV for \MO, respectively. 
    We used standard projector augmented wave pseudopotentials\cite{PhysRevB.50.17953} for \MO, whereas in \TO-compounds we used standard pseudopotentials for Ti (treating d- and s-orbitals as valence) and soft O and F pseudopotentials.
    Effective \textit{U} values for the different materials were set to 10~eV for the $p$-orbitals of O atoms in \MO within the suggested range of \textit{U}-values \cite{falletta_hubbard_2022} and 3.9~eV for the $d$-orbitals of Ti atoms in \TO\ materials as previously determined by constrained RPA \cite{setvin_direct_2014}.
    Polarons were induced by either adding one excess electron for \TO, removing an electron in \MO, or by replacing 1 of the 192 O with an F-dopant in \TOF.

\subsection{Initial Dataset}
    The training dataset for every material was constructed from frames of Molecular Dynamics (MD) runs in the NVT ensemble using Nosé-Hoover thermostat~\cite{Nose1984} with a 1~fs timestep and the computational setup described in the previous section.
    The systems were first thermalized in a shorter run performing a temperature ramping through a velocity rescaling that increased the temperature by 0.5~K at every step until the desired temperature. 
    \MO data were collected at 600~K, \TO and \TOF data at 500~K.
    The temperature was chosen to provide a sufficient number of polaron hopping events within the initial dataset generation.
    In particular, the chosen temperatures resulted in $\sim$20 hopping events for every material after the thermalization.

    The 10 ps FPMD dataset consisting of $10^4$ structures was first divided by reserving the last 20\% of the frames as test set, while the training and validation set were constructed from the inital 80\%.
    To sufficiently represent the possible hopping pathways in the training dataset, we extracted all jumps frames from the remaining 80\%, where we selected 20 frames before and after maximal magnetization changed from one site to another.
    Randomly selected structures and two randomly selected hopping events were used in the validation set.
    The remaining hopping structures in combination with randomly selected structures were used for training.
    For train and validation datasets, a 1 to 2 proportion of hopping to random structures was used.

\subsection{Active learning data generation}
We extract hopping events from long initial MLMD runs performed with a model trained on the initial training dataset to enhance polaron transition sampling.
This process might be unnecessary for simple materials with few symmetrically inequivalent transition pathways (i.e., \MO), but becomes necessary in more complex materials such as \TO.
Polaron hopping events are extracted from the MLMD simulations by identifying timesteps in which $\mathrm{argmax}(\mathbf{m})$ changed.
Similarly to the previous section, we extracted configurations of 40 timesteps centered around the hopping event in MLMD and fed them back into DFT. 
For every series of extracted configurations $\{\mathbf{r}_i\}_{i=1}^{40}$, we performed self consistent computations on all $\mathbf{r}_i$ using the charge density and wavefunction of the previous timestep $\mathbf{r}_{i-1}$, reproducing the operating principle of FPMD.
Additionally, we diversified our sampling by extracting hopping events with distinct transition pathways (\eg\ regarding the crystallographic direction or the relation to other symmetry-breaking features such as dopants) and utilizing hoppings from various temperatures.
Randomly sampled structures from this hopping database were then fed back to the training, validation, and test set to retrain the model.
In cases of multiple symmetry-breaking features (such as in \TOF), extrapolation issues become more severe because many polaronic configurations are not sampled during the initial FPMD run. To address this, we augment the training database with snapshots of transitions to and from each site, as well as random snapshots of the polaron residing at a site for extended periods, ensuring coverage of all distinct polaronic configurations.

\section{Machine learning Methods}
\subsection{Training}
    The model was optimized to fit energies $E$, forces $\mathbf{F}$ and the occupation of each site.
    While fitting energies and forces is a standard procedure to fit potential energy surface models, fitting the occupation requires additional loss terms.
    In particular, we fit the invariant quantity $\Tr[\mathbf{n}^\sigma_i]$, whose difference in spin channels $\sigma=\uparrow,\downarrow$ equal the magnetization of the $i$-th atom.
    For simplicity of notation, we shorten $\Tr[\mathbf{n}^\sigma_i]$ as $n^\sigma_i$, and the difference $n^\uparrow_i - n^\downarrow_i$ as the magnetization $m_i$.
    The total magnetization of a configuration $\sum_i m_i$ is labeled as $M$.
    We found that fitting and predicting $n^\uparrow_i$ and $n^\downarrow_i$ directly resulted in the most stable model.
    Remaining quantities such as the atomic magnetization $m_i$, the total magnetization $M$, or the total atomic occupation $n^\uparrow_i + n^\downarrow_i$ can be derived from these model outputs and are fitted by soft constraints implemented in our loss function.
    By assuming a batch of $K$ structures, each containing $N$ atoms, we can write the loss function $\mathcal{L}$ used to fit the model as follows:
    \begin{align*}
        \mathcal{L} = \frac{1}{K}\sum_{k=1}^K&\left\{ 
        \lambda_E \left(E^\mathrm{ML}_k - E^\mathrm{DFT}_k\right)^2  
        + \frac{\lambda_F}{3N}\sum_{i=1}^{N}\sum_{j=1}^3 (F^\mathrm{ML}_{kij} - F^\mathrm{DFT}_{kij})^2  
        + \lambda_M  \left(M^\mathrm{ML}_{k} - M^\mathrm{DFT}_{k}\right)^2 \right.+\\ 
        & \left.\frac{\lambda_T}{N}\sum_{i=1}^N\left[\sum_{\sigma=\uparrow,\downarrow} (n_{ki}^{\sigma, \mathrm{ML}} - n_{ki}^{\sigma, \mathrm{DFT}})^2 + (m_{ki}^\mathrm{ML} - m_{ki}^\mathrm{DFT})^2 +  \left(\sum_{\sigma=\uparrow,\downarrow}n_{ki}^{\sigma, \mathrm{ML}} - n_{ki}^{\sigma, \mathrm{DFT}}\right)^2\right]\right\}.
    \end{align*}
    
   The weight parameters $\lambda$ are set based on whether they apply to global or atomic quantities.
   Following the same approach as the NequIP paper\cite{batzner_e3-equivariant_2022}, we set the weight parameters of atomic quantities (i.e., $\lambda_F$ for forces and $\lambda_T$ for occupations) to $N^2$, while for global quantities (i.e., $\lambda_E$ for energies and $\lambda_M$ for the total magnetization) the weight is set to 1.
   Similarly to the suggestions of the paper we used a batch size of 1, which we found to produce the lowest fitting errors.
   The model was implemented in JAX\cite{jax2018github} and optimization was performed using the optax-library's implementation of the Adam optimizer\cite{kingma2017} with a learning rate of $5\times 10^{-4}$.
   A decrease on plateau procedure was used during training -- halving the learning rate if fitting errors did not improve in 5 consecutive epochs.
   Using this procedure the model usually converged after $\sim$7 hours of training on an Nvidia Ampere 100 graphic card, requiring on the order of $10^3-10^4$ epochs.

\begin{table}
    \begin{ruledtabular}
        \begin{tabular}{lccccccccc}
            \textrm{RMSE} & 
            \multicolumn{3}{c}{$E$ in meV/Atom} & 
            \multicolumn{3}{c}{$\textbf{F}$ in meV/\AA} & 
            \multicolumn{3}{c}{\textbf{n}} \\
            \colrule
            & \textrm{Train} & \textrm{Validation} & \textrm{Test} & 
              \textrm{Train} & \textrm{Validation} & \textrm{Test} & 
              \textrm{Train} & \textrm{Validation} & \textrm{Test} \\
            \colrule
            \TO+\e & 0.074 & 0.069 & 0.056 & 16.421 & 17.154 & 16.481 & 0.002 & 0.002 & 0.002 \\
            \MO+\h & 0.116 & 0.112 & 0.129 & 20.640 & 26.895 & 25.730 & 0.002 & 0.003 & 0.003 \\
            \TOF   & 0.114 & 0.139 & 0.133 & 17.651 & 19.591 & 15.871 & 0.002 & 0.002 & 0.001 \\
            \colrule
              & 
        \multicolumn{3}{c}{$\textbf{F}_\mathrm{pol}$ in meV/\AA} & 
        \multicolumn{3}{c}{$n^\uparrow_\mathrm{pol} + n^\downarrow_\mathrm{pol}$} & 
        \multicolumn{3}{c}{$m_\mathrm{pol}$} \\
        \colrule
        & \textrm{Train} & \textrm{Validation} & \textrm{Test} & 
          \textrm{Train} & \textrm{Validation} & \textrm{Test} & 
          \textrm{Train} & \textrm{Validation} & \textrm{Test} \\
          \colrule

        \TO+\e & 38.330 & 46.186 & 48.038 & 0.010 & 0.009 & 0.011 & 0.032 & 0.028 & 0.034 \\
        \MO+\h & 29.199 & 45.035 & 44.592 & 0.003 & 0.005 & 0.004 & 0.012 & 0.020 & 0.015 \\
        \TOF   & 46.206 & 65.600 & 40.274 & 0.011 & 0.015 & 0.009 & 0.025 & 0.040 & 0.022 \\
            
        \end{tabular}
    \end{ruledtabular}
\caption{Root mean squared errors of the models for energy $E$, forces $\textbf{F}$ and occupations $\mathbf{n}$ and root mean squared errors of the models at the polaronic site for forces $\textbf{F}_\mathrm{pol}$, the total occupation $n^\uparrow_\mathrm{pol} + n^\downarrow_\mathrm{pol}$, and the magnetization $m_\mathrm{pol}$.}
\label{tab:errors}
\end{table}

Supplementary Table \ref{tab:errors} presents the achieved RMSE for key quantities within our test systems, which are within the expected range for the NequiP architecture \cite{batzner_e3-equivariant_2022}.

\subsection{Machine-learned molecular dynamics}
    The machine-learned molecular dynamics (MLMD) was performed using the JAX-MD framework~\cite{jaxmd2020}.
    To be consistent with the training dataset every simulation was performed in the NVT ensemble using the Nosé-Hoover thermostat and with a timestep to 1~fs.
    Simulations were performed at various temperatures ranging from 100~K to 600~K in steps of 100~K.
    At 100~K polaron hopping was usually too infrequent to perform statistical analysis.
    Stable simulations for the full 10~ns where achieved for \TOF\ and \MO+\h\ at every temperature, while MLMD simulations in \TO+\e\ became unstable for 500~K and 600~K, still allowing to collect statistics for 4~ns and 1.3~ns, respectively.

    \subsection{Hopping at the ML-level}

    \begin{figure}[b!]
        \centering
        \includegraphics[width=0.95\linewidth]{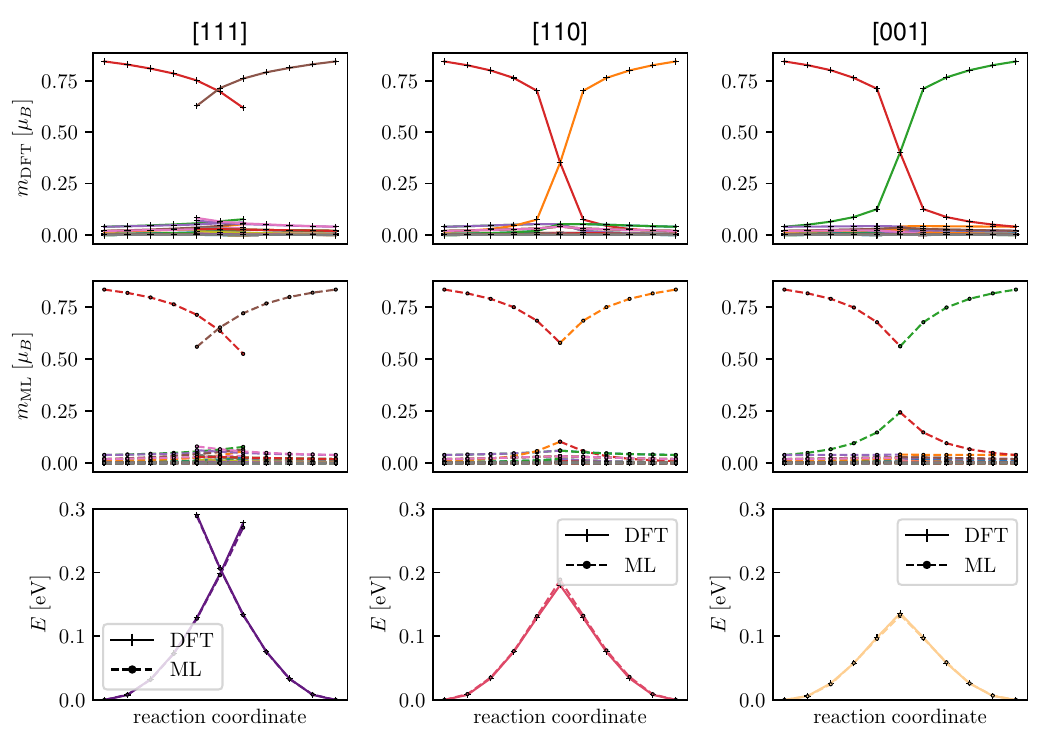}
        \caption{Comparison of DFT- and ML-predicted magnetizations $m$ along different hopping pathways within \TO+\e.}
        \label{fig:transition}
    \end{figure}
    Small polaron dynamics within FPMD is driven by thermally activated hops, where transitions occur in few time steps (tens of fs) of the dynamical simulation.
Most of the time, the polaron remains localized at a specific site. During these intervals, our assumption of an integer charge state holds well, as the polaron can be assigned to a single site. However, during the transition steps of the rare hopping events, this approximation may break down. In these cases, the polaron hopping occurs through the coupling of the initial and final polaronic states, forming an adiabatic transition state where the charge is briefly delocalized across two sites. Additionally, due to thermal distortions, the polaronic band can approach the conduction band, resulting in partial occupations of the conduction band. During electronic minimization, this can lead to the localization of the polaron at a different site.

    In both of these instances, our approximation complicates the explicit treatment of these events using machine learning, as the model is trained under the assumption that the polaron is localized on a single site. Nevertheless, we have found that this assumption produces the most reliable models in terms of long-term stability. The artifacts resulting from this issue are evident in the figures below.
    
    In Supplementary Figure \ref{fig:transition} the magnetization along the reaction coordinate 
    as predicted from DFT and ML for three distinct transition pathways ([001], [110], [111]) in rutile \TO+\e are displayed.
    Reaction coordinates are determined by linear interpolation of the relaxed initial and final polaronic configurations:
    \begin{equation}
        \mathbf{r}_\mathrm{int} = \alpha \mathbf{r}_\mathrm{init} + (1-\alpha) \mathbf{r}_\mathrm{final}
    \end{equation}
It is apparent that for the [110] and [001] transitions, the central interpolated structure (i.e., $\alpha=0.5$) cannot reproduce the desired shared state at the ML level, due to the assumption that the polaronic state is assigned to either site. This is reflected in the ML model, which produces two distinct sets of magnetizations at the central structure, depending on the assignment of the polaronic state. Nevertheless, we found that the model yields well-defined energies for these states, independent of the specific polaronic state. Furthermore, polaronic hopping still occurs during the MLMD, with well-reproducible magnetizations for these adiabatic transitions at the DFT level.

    \begin{figure}
        \centering
        \includegraphics[width=0.95\linewidth]{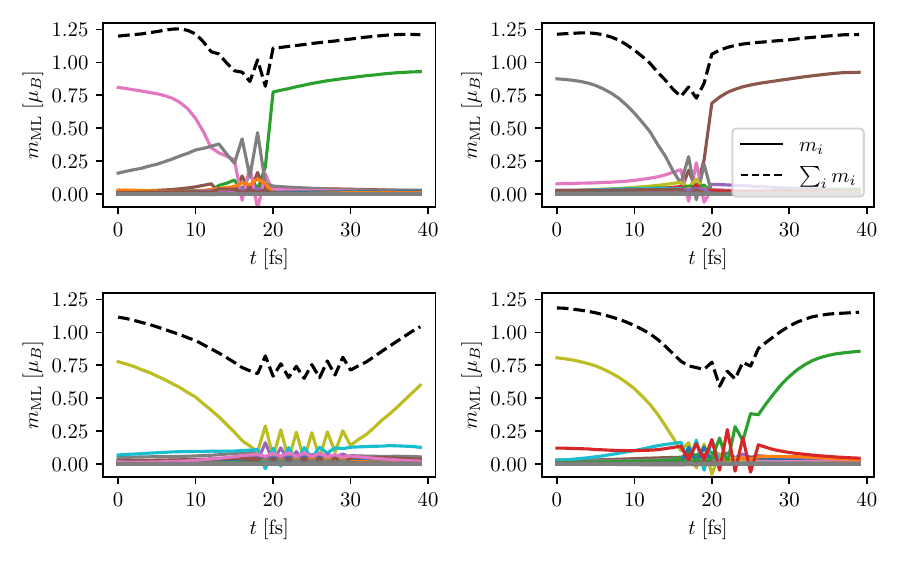}
        \caption{Oscillations in ML-predicted magnetization $m$ in \TO+\e associated to partial delocalization of the polaron.
        The total magnetization $\sum_i m_i$ is shown in dashed black and indicates partial delocalization of the polaron.}
        \label{fig:oscillation}
    \end{figure}
    
    We also found that Leopold can reliably detect partial delocalization in the conduction band.
    This behavior is primarily associated with migrations along [110] and [111]. There, a drop of ML-predicted total magnetization during MLMD is an indicator for small energy differences between the polaronic and conduction band in \TO~at the DFT-level.
    In these cases the model might produce non-physical oscillations as depicted in Supplementary Figure \ref{fig:oscillation}.
    At DFT reference level a partial occupation of the conduction band might result in a favorable initialization of the electronic wavefunction to instantaneously transfer a polaron to a different site.
    In MLMD, we found that explicitly searching for a suitable polaron localization site did not offer any additional benefit.
    The simpler criterion outlined in the maintext in Figure 1b (assigning the polaron to the maximally magnetized site) was found to enhance long term stability of the simulations and reproducibly transfers polarons adiabatically at the DFT-reference level.
    We note that it might be beneficial to include a search for the most stable polaronic sites to mimic the behavior observed in the reference DFT simulations.

\section{Polaron Mobility}

Resulting long MLMD simulations give access to the dynamical properties of the polaron, which we study through the computation of its mean squared displacement $\mathrm{MSD}(t)$, defined as:
\begin{equation}
    \mathrm{MSD}(t) = \left\langle \frac{1}{T - t}\int_0^{T - t} [\mathbf{r}(t + \Delta) - \mathbf{r}(\Delta)]^2 \mathrm{d} \Delta \right\rangle.
\end{equation}
Here, $\langle \cdot \rangle$ represents an ensemble average over all polarons in the system while $\mathbf{r}(t)$ are the polarons trajectories in a simulation with length $T$.
Since only a single polaron is present in our system, the ensemble average in the formula can be omitted, as the dynamics of a single polaron fully characterize the charge transport properties in this case. 
From the obtained MSD we extracted the polaron diffusion coefficient $D$ by using the known relationship
\begin{equation}
    \label{eq:Diff}
    \mathrm{MSD}(t) = 2nDt,
\end{equation}
where $n$ is the dimensionality of the considered trajectory $\mathbf{r}$ (in our case $n = 3$).
Following Eq.~(\ref{eq:Diff}) the diffusion coefficient was extracted by linearly interpolating the MSD function obtained in the linear regime.
Once $D$ has been extracted, we can transform using the Einstein relation $\mu = qD/k_\mathrm{B}T$~\cite{4516} to obtain the final estimate for the polaron mobility.

In this work, we also studied the angular dependence of the mobility in rutile \TOF.
To obtain $\mu(\theta)$ within the (110)-plane containing the dopant, we projected the trajectory $\mathbf{r}(t)$ onto $\hat{\mathbf{n}}(\theta)$ giving access to the direction-dependent trajectory $r_{\theta}(t)$.
We compute the MSD as follows:
\begin{equation}
    \mathrm{MSD}(t, \theta) = \left\langle \frac{1}{T - t}\int_0^{T - t} [r_{\theta}(t + \Delta) - r_{\theta}(\Delta)]^2 \mathrm{d} \Delta \right\rangle\hspace{1cm}\mathrm{with}\hspace{1cm}r_{\theta}(t) = \mathbf{r}(t)\cdot\hat{\mathbf{n}}(\theta).
\end{equation}
In this way we reduce the MSD computation to a 1D trajectory in a certain direction in space, rendering Eq.~(\ref{eq:Diff}) into the form
\begin{equation}
    \mathrm{MSD}(t, \theta) = 2D(\theta)t,
\end{equation}
from which $D(\theta)$ can be extracted and transformed into $\mu(\theta)$ using the same approach described before.

\bibliography{bib}